\documentclass[showpacs,showkeys,superscriptaddress]{revtex4}

\usepackage{amssymb}
\usepackage{amsmath}
\usepackage{graphicx}

\usepackage[english]{babel}
\usepackage[latin1]{inputenc}

\usepackage{amssymb}
\usepackage{enumerate}
\usepackage{amsfonts,amsmath}
\usepackage{bm}
\usepackage{graphicx}
\usepackage{graphics,epsf,psfrag,pifont,dsfont,bbold}
\usepackage{siunitx}
\usepackage{soul,xcolor}
\usepackage{xspace}
\usepackage[normalem]{ulem}
\usepackage{verbatim}
\usepackage{hyperref}

\usepackage{calc}
\newsavebox\CBox
\newcommand\hcancel[2][0.5pt]{%
  \ifmmode\sbox\CBox{$#2$}\else\sbox\CBox{#2}\fi%
  \makebox[0pt][l]{\usebox\CBox}%
  \rule[0.3\ht\CBox-#1/2]{\wd\CBox}{#1}}

\renewcommand\st[1]{{}^{_{\text{#1}}}\!}

\definecolor{dB}{rgb}{.0,.8,.6}
\definecolor{MS}{rgb}{1.,0.,0.}

\newcommand{\be}{\begin{equation}}
\newcommand{\ee}{\end{equation}}
\newcommand{\ben}{\begin{equation*}}
\newcommand{\een}{\end{equation*}}
\newcommand{\pt}{\partial}
\newcommand{\grad}{\bm{\nabla}}
\renewcommand{\d}{\textup{d}}
\newcommand{\Round}[1]{{\left(#1 \right)}}
\newcommand{\Square}[1]{{\left[ #1 \right]}}
\newcommand{\Angle}[1]{{\langle #1  \rangle}}
\newcommand{\bv}[1]{{\bm #1}}

\newcommand{\rv}{\bv{r}}
\newcommand{\cv}{\bv{v}}
\newcommand{\vv}{\bv{U}}
\newcommand{\qv}{\bv{q}}
\newcommand{\av}{\bv{a}}
\newcommand{\jv}{\bv{j}}
\newcommand{\eq}{\st{eq}}
\newcommand\kB{k_{\st{B}}}
\newcommand\A{\st{A}}
\newcommand\B{\st{B}}
\renewcommand\S{\st{S}}

\begin{document}

\title{Lattice Boltzmann Simulations of Non-Equilibrium Fluctuations in a Non-Ideal Binary Mixture}

\author{Daniele Belardinelli}
\author{Mauro Sbragaglia}
\author{Roberto Benzi}
\affiliation{Department of Physics \& INFN, University of Rome ``Tor Vergata'', Via della Ricerca Scientifica 1, 00133, Rome, Italy}
\author{Sergio Ciliberto}
\affiliation{Laboratoire de Physique de Ecole Normale Sup\'erieure de Lyon (CNRS UMR5672), 46 All\'ee d'Italie, 69364, Lyon, France}
\date{\today}
\begin{abstract}
In the recent years the lattice Boltzmann (LB) methodology has been fruitfully extended to include the effects of thermal fluctuations. {So far}, all studied cases pertain equilibrium fluctuations, i.e. fluctuations with respect to an equilibrium background state. In this paper we take a step further and present results of fluctuating LB simulations of a binary mixture confined between two parallel walls in presence of a constant concentration gradient in the wall-to-wall direction. This is a paradigmatic set-up for the study of non-equilibrium (NE) fluctuations, i.e. fluctuations with respect to a non-equilibrium state. {We analyze the dependence of the structure factors for the hydrodynamical fields on the wave vector $\qv$ in both the directions parallel and perpendicular to the walls, as well as the finite-size effects induced by confinement, highlighting the long-range ($\sim |\qv|^{-4}$) nature of correlations in the NE framework.} Results quantitatively agree with the predictions of fluctuating hydrodynamics. Moreover, in presence of a non-ideal (NI) equation of state of the mixture, we also observe that the (spatially homogeneous) average pressure changes, due to a genuinely new contribution triggered by the long-range nature of NE fluctuations. These NE pressure effects are studied at changing the system size and the concentration gradient. Taken all together, we argue that these findings are instrumental to boost the applicability of the fluctuating LB methodology in the framework of NE fluctuations{, possibly in conjunction with experiments.}
\end{abstract}


\maketitle

\section{Introduction}
\begin{figure*}
\includegraphics[width=\textwidth]{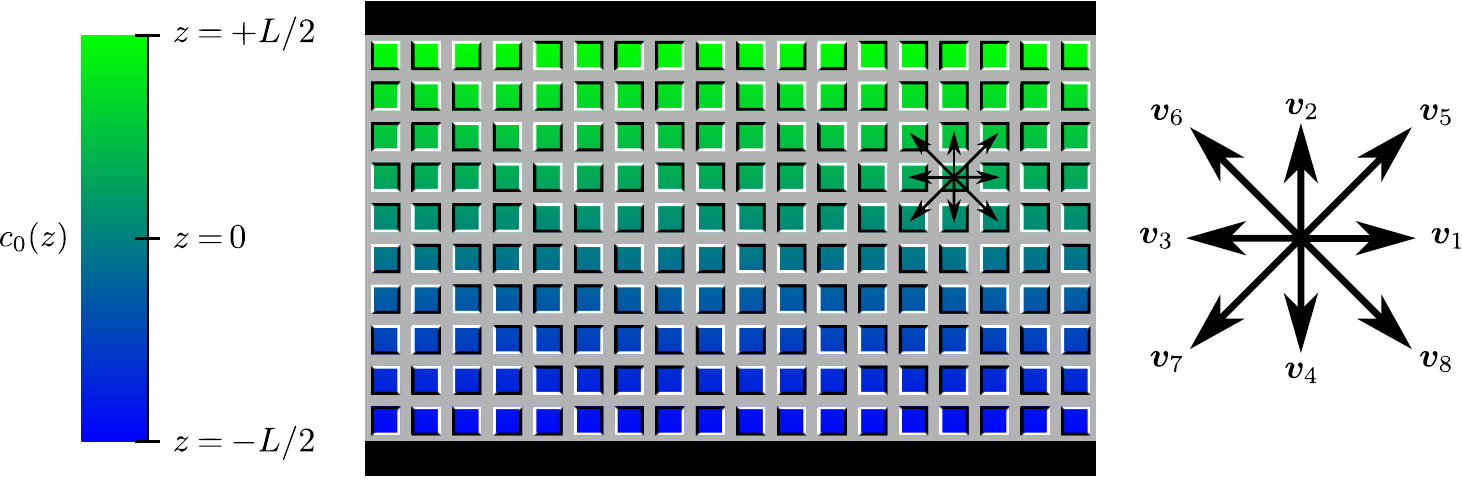}
\caption{{Setup for the numerical simulations. The distance $L$ is the wall-to-wall distance and we take the convention that $z=0$ indicates the center of the channel. A linear concentration background profile $c_0(z)=1/2+z\nabla c_0$ is imposed, corresponding to a constant concentration gradient $\nabla c_0$ in the vertical direction. The vectors $\cv_{1-8}$, together with $\cv_0\equiv\bm{0}$, act as lattice links in the D2Q9 LB simulations.}\label{fig:setup}}
\end{figure*}
The equations of \emph{fluctuating hydrodynamics} supplement the deterministic equations of hydrodynamics with the effect of thermal fluctuations~\cite{Landau}. In a nutshell, the key idea is that whenever scales of observations are small enough, thermal fluctuations cannot be ignored anymore and the non-equilibrium (NE) fluxes in the conservation equations (i.e. diffusion, viscous, etc) need to be promoted to stochastic variables. By linearizing with respect to a homogeneous background and applying the fluctuation dissipation theorem (FDT), one obtains the structure factors for the hydrodynamical fields in agreement with the corresponding statistical mechanics predictions~\cite{Reichl}. Away from criticality, correlations come out to be short-ranged, and the experimental observations with light scattering and neutron scattering techniques confirm such predictions~\cite{BernePecora1976,earnshaw1987,earnshaw1993}. The assumption of full (thermodynamic) equilibrium of the background system greatly simplifies the theoretical approach to the study of thermal fluctuations, but is actually inappropriate in many situations where we have mechanical equilibrium even in presence of temperature or concentration gradients. This may be the case of a Rayleigh-B\'enard cell~\cite{law1990,segre1992} or the case of a binary mixture under the effect of an external field~\cite{vailati1997,croccolo2007,croccolo2016a,croccolo2016b,oprisan2013}. For such systems, the theory of equilibrium thermal fluctuations can be extended~\cite{ZarateBook} to predict fluctuations with respect to a non-equilibrium steady state. In general, NE effects are promoted by two sources: one source is the ``mode'' coupling between the fluctuating velocity and the background inhomogeneous scalar field under consideration, the temperature in single-component fluids~\cite{Kirkpatrick1982b,RonisProcaccia1982,KirkpatrickCohen1983,SchmitzCohen1985,SchmitzCohen1985b,LawSengers1989}, and both the temperature and the concentration for mixtures~\cite{LawNieuwoudt1989,NieuwoudtLaw1990,segre1993}. Another source can be identified in the spatial inhomogeneity of the thermal noise, due to the proportionality of the noise correlations to the temperature~\cite{Proccacia1979,Ronis1980,Tremblay1981,Tremblay1984}, as stated by FDT. Typically, the effects induced by inhomogeneity in the noise are negligible with respect to the mode coupling effect~\cite{ZarateSengers04}. The mode coupling effect causes the small-scale behaviour {of scalar fluctuations} to be divergent as $\sim |\qv|^{-4}$, with $\qv$ being the Fourier wave vector. This was first obtained in~\cite{Kirkpatrick1982b} in the framework of non-equilibrium statistical mechanics and later assessed in the framework of fluctuating hydrodynamics~\cite{RonisProcaccia1982,LawSengers1989}. Experimental confirmations followed~\cite{law1990,segre1992,vailati1997,croccolo2007,oprisan2013,bedeaux2015}. We emphasize that NE fluctuations cause long-range correlation effects; similar long-range correlation effects are absent in equilibrium situations, except close to criticality~\cite{gambassi2009}. Moreover, the long-range nature of the NE effects causes fluctuation-induced forces. This feature has been extensively discussed in the literature~\cite{Kirkpatrick13,Kirkpatrick14,Kirkpatrick15,kirkpatrick2015,kirkpatrick2016,kirkpatrick2016b}. Similar NE pressure effects have also been studied in the non-linear Navier-Stokes equations with imposed shear rate~\cite{lutsko1985,wada2003,Zarate2008,varghese2017}, where the NE effects are triggered by the non linear coupling between the imposed shear rate {and} the flow itself.\\ 
Thermal fluctuations become relevant at mesoscales, where many complex hydrodynamic phenomena occur, like for example the motion of non-ideal (NI) interfaces~\cite{aarts2004}, the coupling between colloidal particles and the fluid~\cite{lipowsky2005,einert2005}, the rheology of vesicles and red blood cells~\cite{noguchi2005,park2010,fedosov2010}, the acoustic-magnetic effect in magnetic fluids~\cite{storozhenko2011}. The need of understanding complex hydrodynamic phenomena at mesoscales naturally sets a compelling case for the development of suitably designed numerical methods. Beyond the numerical simulations based on the continuum equations of hydrodynamics~\cite{defabritiis2007,balboa2012}, in the recent years mesoscale simulations based on the \emph{lattice Boltzmann} (LB)~\cite{Benzi92,ChenDoolen98} have been proposed~\cite{Varnik11,Dunweg,KaehlerWagner13}. The LB method stands out due to its remarkable capability of handling complex boundary conditions and NI fluids with phase transitions/segregation~\cite{DunwegReview,Aidun10,Zhang11,Kangetal14,Schilleretal18}; hence the LB coupled with thermal fluctuations is a promising pathway for realizing very powerful mesoscale simulation methods. The idea of including noise in LB, in fact, has constituted an active research field of the recent years~\cite{Ladd,Adhikari,Dunweg,Gross10,Gross11,KaehlerWagner13,Gross10,Gross11,belardinelli2015}. All these implementations, however, consider hydrodynamical systems fluctuating around a state in full equilibrium. The aim of the present paper is to explore the applicability of the fluctuating LB in the context of NE fluctuations. While none of the approaches proposed in the literature~\cite{Ladd,Adhikari,Dunweg,Gross10,Gross11,KaehlerWagner13,Gross10,Gross11,belardinelli2015} can be trivially extended to the case with temperature gradients in the background, in~\cite{belardinelli2015} it is discussed how to correctly formulate noise in multicomponent systems, even in presence of an inhomogeneous background concentration $c_0(\rv)$. Numerical simulations showed convincing agreement between the numerically evaluated equilibrium structure factors and the theoretical predictions. The latter, which can be obtained directly in the kinetic framework (see~\cite{belardinelli2015}), coincide with the predictions of fluctuating hydrodynamics. However, this is obviously not enough to prove convergence of fluctuating LB towards fluctuating hydrodynamics. Indeed, the stochastic noise terms break one of the basic assumptions of Chapman-Enskog theory (i.e. having fields slowly varying in space and time). Hence, the coincidence of theoretical results (kinetic framework {\it vs.} hydrodynamics framework) seems like a lucky case, possibly valid in homogeneous cases. Hence, investigating NE in LB simulations, is a further way to highlight the convergence of fluctuating LB towards fluctuating hydrodynamics. The article is organized as follow. In section \ref{sec:System} the system and its governing equations are presented. The used methodology is described in section \ref{sec:Methodology}. The numerical results, both in equilibrium and out of equilibrium, are discussed and compared with the theoretical predictions in section \ref{sec:RESULTSandDISCUSSION}. We conclude in section \ref{sec:Conclusions}. The appendix recalls some relevant definitions.
\section{System\label{sec:System}}
In this paper we study the problem of NE fluctuations by considering a two dimensional binary mixture confined between two walls in presence of a constant concentration gradient $\nabla c_0$ in the wall-to-wall direction (see Figure~\ref{fig:setup}). The reference fluctuating hydrodynamical equations for the concentration and velocity fluctuations $\delta c=c-c_0$ and $\delta\vv=(U_x,U_z)$ are (see~\cite{ZarateBook} and references therein) 
\be\label{eq:linearized1}
\grad \cdot \delta \vv = 0,
\ee
\be\label{eq:linearized2}
\bar{\rho}(\pt_t\delta c + U_z\nabla c_0) = \bar{\rho} D\nabla^2\delta c - \grad\cdot{\bm J},
\ee
\be\label{eq:linearized3}
\bar{\rho}\pt_t\delta\vv = \bar{\rho}\nu\nabla^2 \delta\vv - \grad\cdot{\bm \Pi},
\ee
where $\bar{\rho}$, $D$ and $\nu$ are reference values for total mass density, mass diffusion coefficient and kinematic viscosity, respectively. The terms ${\bm J}$ and ${\bm \Pi}$ are the stochastic contributions to the deterministic equations of hydrodynamics. Specifically, ${\bm J}$ is a stochastic flux and ${\bm \Pi}$ is a stochastic stress tensor satisfying FDT:
\begin{equation}\label{eq:noiseJ}
\Angle{J_i(\rv,t) J_j(\rv',t')}=2 \kB T \bar{\rho} D \chi \delta(\rv-\rv')\delta(t-t'),
\end{equation}
\begin{equation}\label{eq:noisePi}
\Angle{\Pi_{ij}(\rv,t) \Pi_{kl}(\rv',t')} =2 \kB T \bar{\rho}\nu \Delta_{ijkl} \delta(\rv-\rv')\delta(t-t'),
\end{equation}
with $\Delta_{ijkl}=\delta_{ik}\delta_{jl}+\delta_{il}\delta_{jk}$, $\kB$ the Boltzmann constant and $T$ the temperature, while $\chi$ indicates the inverse osmotic susceptibility: $\chi^{-1}=\left(\pt \mu/\pt c\right)_{P,T}$, with $\mu$ the chemical potential and $P$ the fluid pressure.\\
\section{Methodology\label{sec:Methodology}}
%
\begin{table}[t!]
\begin{center}
\begin{tabular}{| r | c | c | c | c |}
\hline
& & & & \\
$a$ & $T_{a l}$ & $m_a$ & $m^{\eq}_a(\rho,\vv)$ & $\lambda_a$\\
\hline
 0 & 1                                    & $\rho$          & $\rho$ & $\lambda_0$\\
 1 & $(\cv_l)_x$                          & $j_x$         & $\rho U_x$ & $\lambda_{\st{d}}$\\
 2 & $(\cv_l)_z$                          & $j_z$         & $\rho U_z$ & $\lambda_{\st{d}}$\\
\hline
 3 & $3 |\cv_l|^{2} - 2$                  & $e$             & $3 \rho|\vv|^2$ & $\lambda_{e}$\\
 4 & $(\cv_l)_x^2 - (\cv_l)_z^2$        	& ${\Sigma_{w w}}$       & $\rho(U_x^{2} - U_z^2)$ & $\lambda_{\st{s}}$\\
 5 & $(\cv_l)_x(\cv_l)_z$                   & ${\Sigma_{x z}}$       & $\rho U_x U_z$ & $\lambda_{\st{s}}$\\
\hline
 6 & $(3 |\cv_l|^{2} - 4) (\cv_l)_x$      & $Q_x$         & $0$ & $\lambda_{Q}$\\
 7 & $(3 |\cv_l|^{2} - 4) (\cv_l)_z$      & $Q_z$         & $0$ & $\lambda_{Q}$\\
 8 & $9 |\cv_l|^{4} - 15 |\cv_l|^{2} + 2$ & $\epsilon$      & $0$ & $\lambda_{\epsilon}$\\
\hline
\end{tabular}
\end{center}
\caption{Moments set for the D2Q9 model used in the LB simulations. The index S of the species has been omitted. As the set of velocities is finite, the set of $T_{a l}$ forms a basis. The moments $m_a$ are computed according to Eq.~\eqref{eq:modes-discrete}. They relax toward their respective asymptotic values according to a time scale $1/\lambda_a$. The first three {rows} cover the conserved moments.}
\label{tab:mom}
\end{table}
The basic idea behind the LB methodology is to derive the equations of hydrodynamics from the more fundamental kinetic theory~\cite{Shan06}. In this section we briefly review the fluctuating multicomponent LB model that we use. Extensive details are reported in~\cite{belardinelli2015}. The model does not consider directly the hydrodynamic fields, but considers a kinetic description of a multicomponent fluid with two species, say A and B, having mass densities $\rho^{\A}$ and $\rho^{\B}$. The corresponding total mass density is indicated with $\rho=\rho^{\A}+\rho^{\B}$, while mass concentration is conventionally taken as $c=\rho^{\A}/\rho$.\\
The LB method makes use of a set of $Q$ distribution functions $f^{\S}_l(\rv,t)$ ($l=0,\dots,Q-1$), representing the number of particles of the species S $=$ A,B at time $t$ in an elementary lattice cell of unit volume around the position $\rv$ moving with velocity $\cv_l$. Mass densities are recovered as $\rho^{\S}=\sum_l f^{\S}_l$ \footnote{For simplicity, we will neglect differences in molecular masses between the two species by setting each of them equal to unit.}. One then introduces the (isotropic) lattice spacing $\Delta r$ and the time interval $\Delta t$ to rescale positions and times, respectively. Coherently, velocities are rescaled by $\Delta r/\Delta t$. Dimensionless variables will be noted in the same way as the variables themselves. In this way, while $t$ varies on the natural set, the velocities $\cv_l$ act as links connecting the lattice points $\rv$. The LB evolution is described by the following algorithm:
\be\label{eq:LB}
f^{\S}_l(\rv+\cv_l,t+1) = f^{\S}_l(\rv,t) + R^{\S}_l(\rv,t).
\ee
Here, $R^{\S}_l$ is the responsible for the change of $f^{\S}_l$ when moving along the link $\cv_l$ in a time step. It is better written in terms of the moments $m^{\S}_a$ ($a=0,\dots,Q-1$), which are defined by the following invertible transformation~\cite{belardinelli2015}:
\begin{align}\label{eq:modes-discrete}
&m^{\S}_a = \sum_l T_{a l} f^{\S}_l, &&f^{\S}_l = w_l \sum_a \frac{T_{a l}}{N_a} m^{\S}_a.
\end{align}
In table \ref{tab:mom} it is reported the chosen set of $T_{a l}$ for the D2Q9 lattice used in the simulations. This is a 2-dimensional lattice with $Q=9$ velocities (see Figure~\ref{fig:setup}): $\cv_0=(0,0)$, $\cv_{1}=(1,0)=-\cv_{3}$, $\cv_{2}=(0,1)=-\cv_{4}$, $\cv_{5}=(1,1)=-\cv_{7}$, $\cv_{6}=(-1,1)=-\cv_{8}$,
the associated weights being $w_0=4/9$, $w_{1-4}=1/9$ and $w_{5-8}=1/36$. The normalization constants are obtained as $N_a=\sum_l w_l T_{a l}^2$. In particular, lattice mass and momentum densities are given by
\be
\rho^{\S} = m^{\S}_0 = \sum_l f^{\S}_l
\ee
and
\be
\jv^{\S} = (j^{\S}_x,j^{\S}_z) = (m^{\S}_1,m^{\S}_2) = \sum_l \cv_l f^{\S}_l,
\ee
respectively. While the lattice mass densities coincide with their physical counterpart, the physical baricentric velocity $\vv$ is constructed as~\cite{belardinelli2015}
\be\label{eq:bar}
\rho\vv=(\rho U_x,\rho U_z)=\jv^{\A}+\jv^{\B}+\frac{1}{2}\rho\av.
\ee
The additional term {is a lattice correction and it} involves the effective body-force density $\rho\av$ acting on the fluid. We can write $\rho\av=\rho^{\A}\av^{\A}+\rho^{\B}\av^{\B}$, and decompose each term in the sum of non-ideal (NI) and non-equilibrium (NE) contributions by writing for each species $\av^{\S}=\av^{\S}_{\st{NI}}+\av^{\S}_{\st{NE}}$. The former is constructed on the lattice and takes the form~\cite{shan1993lattice,shan1994simulation,bastea2002hydrodynamics,benzi2009mesoscopic,sbragaglia2013interaction}
\be\label{eq:aNI}
\av^{\A}_{\st{NI}}(\rv,t) = - G \sum_l w_l\cv_l\rho^{\B}(\rv+\cv_l,t),
\ee
and an analogous expression holds for $\av^{\B}_{\st{NI}}$, having B replaced by A on the rhs. The positive constant $G$ is the same for both the species and is a tunable parameter in the model~\cite{Sbragaglia07a}. It regulates the intensity of interactions between the two fluids, which are assumed to be separately ideal (in the expression of $\av^{\A}_{\st{NI}}$ only $\rho^{\B}$ appears). {This produces a NI contribution in the equation of state (see Eq.~\eqref{eq:EQUATIONofSTATE} below).} The NE contribution is chosen in such a way that it imposes a concentration gradient $\grad c_0=(0,\nabla c_0)$ in the steady state, which is important for the study of NE effects:
\be\label{eq:ne-acc}
\av^{\A}_{\st{NE}}={\frac{1}{3}}(0,\nabla c_0/c),
\ee
where the prefactor has been conveniently chosen equal to the lattice speed of sound for the D2Q9 model, that is $1/3$~\cite{Shan06}. The analogous expression for $\av^{\B}_{\st{NE}}$ is obtained by replacing $\nabla c_0/c$ with $-\nabla c_0/(1-c)=\nabla c_0/(c-1)$. Notice that the momentum balance {and consequently the pressure are} unaffected by the NE forcing, since
\be\label{eq:NEcond}
\rho^{\A}\av^{\A}_{\st{NE}}+\rho^{\B}\av^{\B}_{\st{NE}} = \bm{0}.
\ee
In this way, the NE acceleration \eqref{eq:ne-acc} gives a contribution in the diffusion current proportional to $\rho\grad c_0$ (see Eq.~\eqref{eq:hydro2} below), thus fixing $\grad c=\grad c_0$ in the stationary steady state \footnote{To allow Eq.~\eqref{eq:hydro2} to be recovered for any value of $\nabla c_0$ a further {contribution} to the acceleration must be added in the form $\av^{\A}_*=(1-c)(\av^{\B}_{\st{NI}}-\av^{\A}_{\st{NI}})$, which satisfies the relation~\eqref{eq:NEcond}, thus giving the same equation of state~\eqref{eq:EQUATIONofSTATE}.}. Moments for $a=3,\dots,5$ are related to transport phenomena, while higher order moments have no hydrodynamical counterpart and constitute the so-called ``ghost'' sector (see table \ref{tab:mom}). Close to a \emph{local} equilibrium state only the first moments contribute, as we can write $m^{\S}_a=m^{\eq}_a(\rho^{\S},\vv)$, with the equilibrium hydrodynamical moments $m^{\eq}_a(\rho,\vv)$ given in table \ref{tab:mom}. With these ingredients, the last term in \eqref{eq:LB} can be written as
\be
R^{\S}_l = w_l \sum_a \frac{T_{a l}}{N_a} (C^{\S}_a+F^{\S}_a+\xi^{\S}_a).
\ee
The first term in the round brackets models the relaxation towards the \emph{local} equilibrium:
\be\label{eq:MRT}
C^{\S}_a = \lambda^{\S}_a[m^{\eq}_a(\rho^{\S},\vv) - m^{\S}_a].
\ee
The dimensionless constants $\lambda^{\S}_a$ are the lattice relaxation frequencies, $1/\lambda^{\S}_a$ being the corresponding lattice relaxation times. This is the multiple relaxation times (MRT) generalization of the celebrated BGK (for Bhatnagar, Gross and Krook~\cite{bhatnagar1954model}) form of the Boltzmann collision integral. All the $\lambda^{\S}_a$ are tunable parameters of the model, with some restrictions imposed by the request of mass and momentum conservations. Since $m^{\eq}_0(\rho^{\S},\vv)=m^{\S}_0=\rho^{\S}$, mass conservation is ensured for each species separately, independently on the actual value of $\lambda^{\S}_0$. Instead, the second argument of the equilibrium distribution in \eqref{eq:MRT} is the baricentric velocity $\vv$ of equation \eqref{eq:bar}, allowing in this way the diffusion {of} one species into the other. Conservation of \emph{total} momentum is then enforced by conveniently choosing
\be
\lambda^{\S}_{1,2}=\lambda_{\st{d}},
\ee
where the lattice diffusion relaxation frequency $\lambda_{\st{d}}$ is the same for both the species. Similarly, the lattice relaxation frequencies associated to the shear moments ${\Sigma_{w w}}$ and ${\Sigma_{x z}}$ ($a=4,5$, see table \ref{tab:mom}) are chosen as
\be
\lambda^{\S}_{4,5}=\lambda_{\st{s}},
\ee
with the lattice shear relaxation frequency $\lambda_{\st{s}}$ being the same for both the species.\\ 
The second term in the round brackets of equation \eqref{eq:LB} models the action of the long-range interactions between the fluid particles. {The first order moments ($a=1,2$) are given by
\be
(F^{\S}_1,F^{\S}_2)=\rho^{\S}\av^{\S}.
\ee
We omit the expression of the moments of order higher than one, for shortness, by remarking that they must be included for a proper simulation of a non-homogeneous fluctuating system~\cite{belardinelli2015}.}\\ 
The last term in the round brackets of \eqref{eq:LB} {accounts for thermal} fluctuations. {These are modelled with} zero-mean Gaussian random variable{s}, uncorrelated in time and with constant covariances (which can however depend on $\rv$). The derivation of the precise expression of the noise covariances has been achieved in~\cite{belardinelli2015}. It makes use of the FDT directly applied at the kinetic level. The covariance matrix appears to be diagonal in both moments \footnote{{For the sake of precision, we mention that off-diagonal noise correlations emerge upon discretization of the velocity space~\cite{belardinelli2015}, which are however negligible for practical purposes.}} and space, as well as in time by construction, allowing us to write
\be
\Angle{\xi^{\S}_a(\rv,t) \xi^{\S'}_{a'}(\rv',t')} = \Angle{\xi^{\S}\xi^{\S'}}_a\delta_{a,a'}\delta_{\rv,\rv'}\delta_{t,t'}.
\ee
The quantities $\xi^{\S}_a$ are arranged in the same way as the moments $m^{\S}_a$. 
In particular,
\begin{align}
&\rho^{\S}_\xi = \xi^{\S}_0, &&\jv^{\S}_{\xi} = (\xi^{\S}_1,\xi^{\S}_2).
\end{align}
As a direct consequence of mass conservation for each species it results that
\be
\Angle{\xi^{\S}\xi^{\S'}}_0=0,
\ee
coherently with {an} identically vanishing $\rho^{\S}_\xi$. Momentum, instead, is not conserved separately for each species, due to diffusion effects. However, total momentum in conserved, so that $\jv^{\A}_{\xi}+\jv^{\B}_{\xi}$ must be identically vanishing. Coherently, it is found
\be\label{eq:noise-bin-expl}
\Angle{\xi^{\S}\xi^{\S}}_{1,2} = -\Angle{\xi^{\A}\xi^{\B}}_{1,2} = (2-\lambda_{\st{d}})\lambda_{\st{d}}\kB T\frac{\rho^{\A}\rho^{\B}}{\rho}.
\ee
Higher order noise correlations have also to be taken into account. The only non vanishing are for S $=$ S$'$:
\be\label{eq:noise-bin-expl-high}
\Angle{\xi^{\S}\xi^{\S}}_a = {3}(2-\lambda^{\S}_a) \lambda^{\S}_aN_a \kB T \rho^{\S} \textup{ for }a=3,\dots,8.
\ee
The factors $\Angle{\xi^{\S}\xi^{\S'}}_a$ would depend on space through their dependence on the background fields $\rho^{\S}$. However, in order to focus on the mode coupling effects, which are dominant for the case at hand~\cite{ZarateSengers04}, we mainly performed simulations by keeping the $\rho^{\S}$ in \eqref{eq:noise-bin-expl}-\eqref{eq:noise-bin-expl-high} equal to their reference values. The effect of inhomogeneities in the $\Angle{\xi^{\S}\xi^{\S'}}_a$ will be highlighted only preliminarily in this study. Of central role in the derivation of the previous noise covariances are the proprieties of the \emph{noiseless} ($\xi^{\S}_a\equiv0$) stationary state reached by the system. This is assumed to have the \emph{local} equilibrium form $m^{\S}_a=m^{\eq}_a(\rho^{\S},\bm{0})=\rho^{\S}\delta_{a,0}$, so that $C^{\S}_a=0$. Thus, by summing \eqref{eq:LB} over the species, using \eqref{eq:NEcond} and \eqref{eq:aNI}, and performing the continuum limit (formally, $\cv_l\to\bm{0}$), we get $\grad\rho = - G \grad(\rho^{\A}\rho^{\B})$. This can be written in the form $\grad P=\bm{0}$, allowing us to deduce the equation of state $P=P(\rho,c)$ for the system at hand:
\be\label{eq:EQUATIONofSTATE}
\begin{split}P &= {\frac{1}{3}} (\rho^{\A} + \rho^{\B}) + {\frac{1}{3}} G\rho^{\A} \rho^{\B}\\
&= {\frac{1}{3}} \rho + {\frac{1}{3}} G\rho^2 c(1-c).\end{split}
\ee
{The ideal equation of state $P=\rho/3$ is recovered by setting $G=0$; recall that the factor $1/3$ equals the D2Q9 lattice speed of sound~\cite{Shan06}.} By applying the Chapman-Enskog procedure and treating the stochastic terms as generic external forces, one can prove~\cite{SegaSbragaglia13} that the fluctuating hydrodynamic equations of a binary mixture with total mass density $\rho$, baricentric velocity $\vv$ and mass concentration $c$ are recovered 
(the superscript $^\intercal$ denotes transposition):
\be\label{eq:hydro1}
\pt_{t} \rho + \grad \cdot (\rho \vv) = 0,
\ee
\be\label{eq:hydro2}
\rho(\pt_{t}c + \vv \cdot \grad c) = \grad \cdot [\rho D \grad (c-c_0) - {\bm J}],
\ee
\be\label{eq:hydro3}
\rho(\pt_{t}\vv + \vv \cdot \grad \vv) = - \grad P + \grad \cdot [\rho\nu (\grad \vv + \grad \vv^\intercal) - {\bm \Pi}],
\ee
where ${\bm J}$ and ${\bm \Pi}$ are the noise fields whose variances are fixed by the Chapman-Enskog procedure, and thus satisfying FDT at kinetic level. The mass diffusion coefficient
\be\label{eq:diff}
D={\frac{1}{3}} \Round{\frac{1}{\lambda_{\st{d}}}-\frac{1}{2}}
\ee
and the kinematic viscosity
\be\label{eq:vis}
\nu={\frac{1}{3}} \Round{\frac{1}{\lambda_{\st{s}}}-\frac{1}{2}}
\ee
respectively regulate the intensity of the diffusion fluxes and the viscous stresses~\cite{SegaSbragaglia13} and are tunable in the model, by specifying $\lambda_{\st{d}}$ and $\lambda_{\st{s}}$ independently. Notice that the total mass density here is a dynamical variable. However, compressibility effects result to be small (see figure \ref{fig:densityprofile}), hence by linearizing the equations around the background state one ends up with Eqs.~\eqref{eq:linearized1}-\eqref{eq:linearized3}. \\
Summarizing, we use the LB solver described in~\cite{belardinelli2015} to simulate the hydrodynamical equations of a binary mixture in presence of a background stationary concentration gradient. If we trust the hydrodynamical limit of the LB model, we can then assess the properties of fluctuations by changing the background gradient $\nabla c_0$, the geometry used, the transport coefficients $D$ and $\nu$, and the interaction strength $G$ that regulates the NI character of the mixture. We again remark that the fluctuating terms violate one of the basic assumptions of Chapman-Enskog theory (i.e. having fields slowly varying in space and time). {We} can only formally obtain Eqs. \eqref{eq:hydro1}-\eqref{eq:hydro3}. Rather, the convergence towards the fluctuating hydrodynamical equations must be assessed via numerical simulations and a careful comparisons with the predictions of fluctuating hydrodynamics~\cite{Kirkpatrick15,ZarateSengers01,ZarateSengers02}.
\subsection{Set-up}
We consider a two dimensional system with dimensions $L_x \times L$, with periodic boundary conditions in the stream-flow ($x$) direction and two solid walls located at $z=\pm L/2$. The two dimensional choice is done to make the many computations affordable at changing $L$ up to few tens of grid points. Indeed, the solutions of fluctuating hydrodynamics assume infinitely long parallel walls~\cite{ZarateBook}; hence, for a given $L$, the stream-flow lengthscale $L_x$ needs to be large enough to prevent spurious effects induced by periodicity \footnote{The choice $L_x=4L$ is enough to obtain negligible spurious effects for wavenumbers as small as $1/L$.}. Regarding the boundary conditions, we choose the mid-way bounce back rule for the LB kinetic populations~\cite{Succi01}: apart from small discrete effects, these provide a no-slip boundary condition for the tangential velocity ($U_x=0$) in absence of fluctuations. We also enforced exactly a zero normal velocity at the wall ($U_z=0$) at every time-step by properly readjusting the rest population at the wall. Regarding the boundary conditions on the concentration field, when computing the NI forces we impose that the densities of both components at the wall are equal to the neighboring fluid nodes~\cite{sbragaglia2008}. Both the no-slip boundary condition and the conditions on the species densities (hence the boundary condition on concentration) are obviously changed by thermal fluctuations. To the best of the authors' knowledge there is no systematic study on the effects induced by thermal fluctuations on the LB boundary conditions and their hydrodynamic manifestations. A systematic study is surely warranted for the future. However, for the purposes of the present paper, we remark that boundary conditions affect the NE spectra only at large scales~\cite{Kirkpatrick15,ZarateSengers01,ZarateSengers02}, while the small-scale behaviour is rather independent of the boundary conditions used. Moreover, regarding the large-scales, there are various solutions of fluctuating hydrodynamics that report the effects of hydrodynamic boundary conditions~\cite{Kirkpatrick15,ZarateSengers01,ZarateSengers02}. Thus, if from one side we can assess the universality in the small-scale behaviour, as a bonus we can also explore preliminarily the robustness of the LB boundary conditions used by direct comparisons against analytical solutions available.\\
From now on, when writing $\bar{\rho}$ we will mean that reference value for the total mass density such that, for given values of $L_x$ and $L$, the product $\bar{\rho} L_x L$ gives the total mass, which is exactly conserved by the algorithm. Furthermore, all the simulations are performed in such a way that $\bar{c}_0\equiv c_0(z=0)=1/2$, and we take $\bar{c}_0$ as the reference value for the concentration. All the numerical results discussed in the following sections will be reported in LB units. In particular, in those units we set $\bar{\rho}=1$.

\section{Results and discussions\label{sec:RESULTSandDISCUSSION}}
\subsection{Equilibrium Fluctuations ($\nabla c_0=0$)\label{sec:EQ}}
\begin{figure}[h!]
\includegraphics[width=.6\columnwidth]{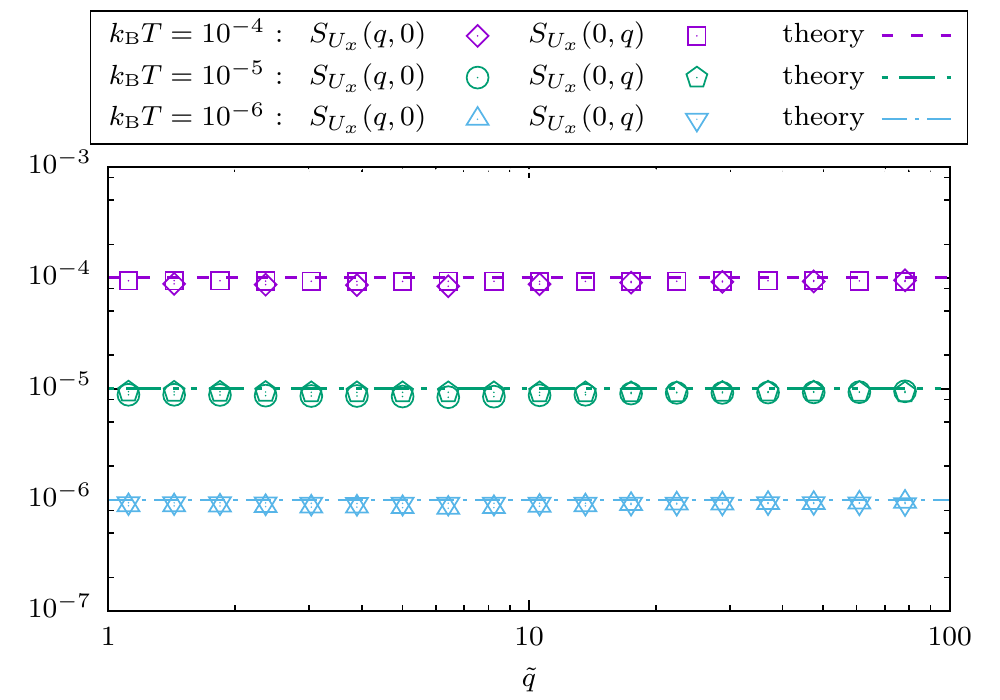}\\
\includegraphics[width=.6\columnwidth]{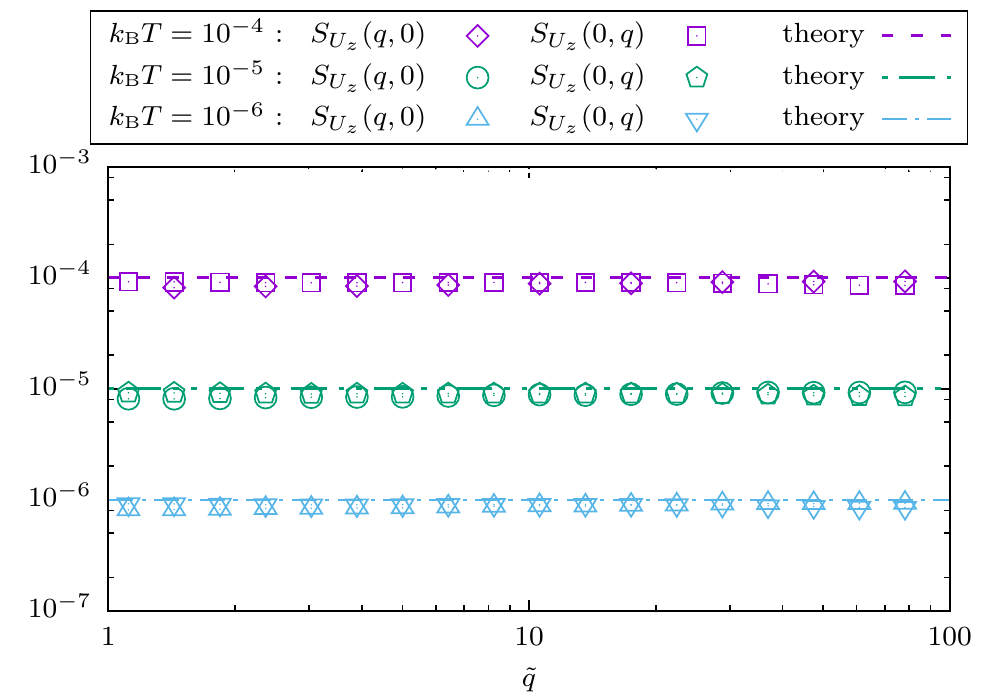}\\
\includegraphics[width=.6\columnwidth]{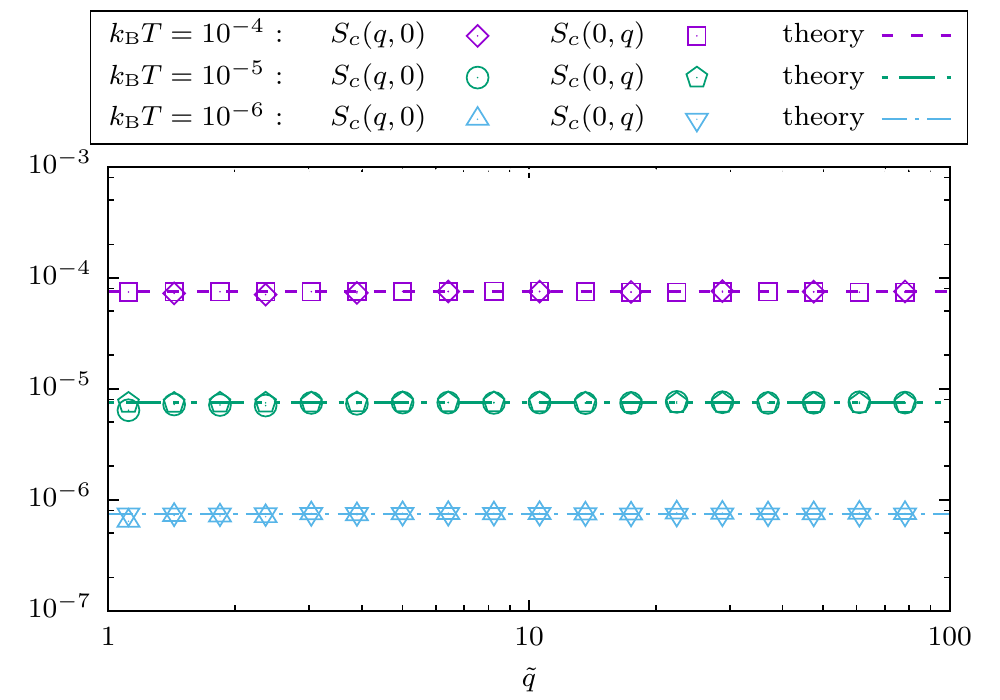}
\caption{Spectra of fluctuations of hydrodynamical fields around an equilibrium background ($\nabla c_0=0$). {\bf Top panel:} Velocity in the streamflow ($x$) direction; theoretical prediction in Eq.~\eqref{eq:VELOCITY_EQ}. {\bf Central panel:} Velocity in the wall-to-wall ($z$) direction; theoretical prediction in Eq.~\eqref{eq:VELOCITY_EQ}. {\bf Bottom panel:} Concentration fluctuations; theoretical prediction in Eq.~\eqref{eq:CONCENTRATION_EQ}.\label{fig:spectrum}}
\end{figure}

The model that we {use} has already been extensively validated in unconfined homogeneous situations in~\cite{belardinelli2015}. However, since we will use confined simulations with wall boundary conditions for the NE fluctuations ($\nabla c_0 \neq 0$), it is mandatory a preliminary characterization of the equilibrium fluctuations ($\nabla c_0=0$) in such confined situations. {These can be studied} in Fourier space through the structure factors of the velocity and concentration fluctuations, respectively $S_{U_{x,z}}(\qv)$ and $S_c(\qv)$ (see Appendix), where
\be
\qv=(q_x,q_z)
\ee
is the wave vector. It is well known~\cite{ZarateBook} that for equilibrium fluctuations the structure factors of both velocity and concentration fluctuations are independent of the wave vector. More quantitatively,
\be\label{eq:VELOCITY_EQ}
S_{U_{x,z}}(\qv)=\Angle{|U_{x,z}(\qv)|^2}=\frac{\kB T}{\bar{\rho}},
\ee
\be\label{eq:CONCENTRATION_EQ}
S_c(\qv)=\Angle{|\delta c(\qv)|^2}={3}\frac{\kB T}{\bar{\rho}}\bar{c}_0(1-\bar{c}_0).
\ee
We thus considered a {\it homogeneous} system ($\nabla c_0=0$) confined in a channel with resolution $L=32${, and correspondingly $L_x=4L=128$}. We then performed simulations at changing $\kB T$ in the range $10^{-6}$--$10^{-4}$. The measured structure factors appear in good agreement with the previous constant values in both the streamflow ($\qv=(q,0)$) and the wall-to-wall ($\qv=(0,q)$) directions, as shown in Figure \ref{fig:spectrum} on varying dimensionless wavenumbers
\be\label{eq:qtilde}
\tilde{q} = q L.
\ee
This corresponds to delta-like correlations in real space:
\be\label{eq:VELOCITY_CORR}
\Angle{U_{x,z}(z)U_{x,z}(0)} = \frac{\kB T}{\bar{\rho}}\delta_{z,0},
\ee
\be\label{eq:CONCENTRATION_CORR}
\Angle{\delta c(z)\delta c(0)} = {3}\frac{\kB T}{\bar{\rho}}\bar{c}_0(1-\bar{c}_0)\delta_{z,0}.
\ee
As shown in the following section, these properties are maintained by the velocities even in presence of {a non zero} concentration gradient, while the concentration itself exhibits long-range spatial correlations (see Figure \ref{fig:correlationrealspace}).
%
\subsection{Non-equilibrium {fluctuations} ($\nabla c_0\neq0$)\label{sec:NEQ}}
\begin{figure}
\includegraphics[width=.6\columnwidth]{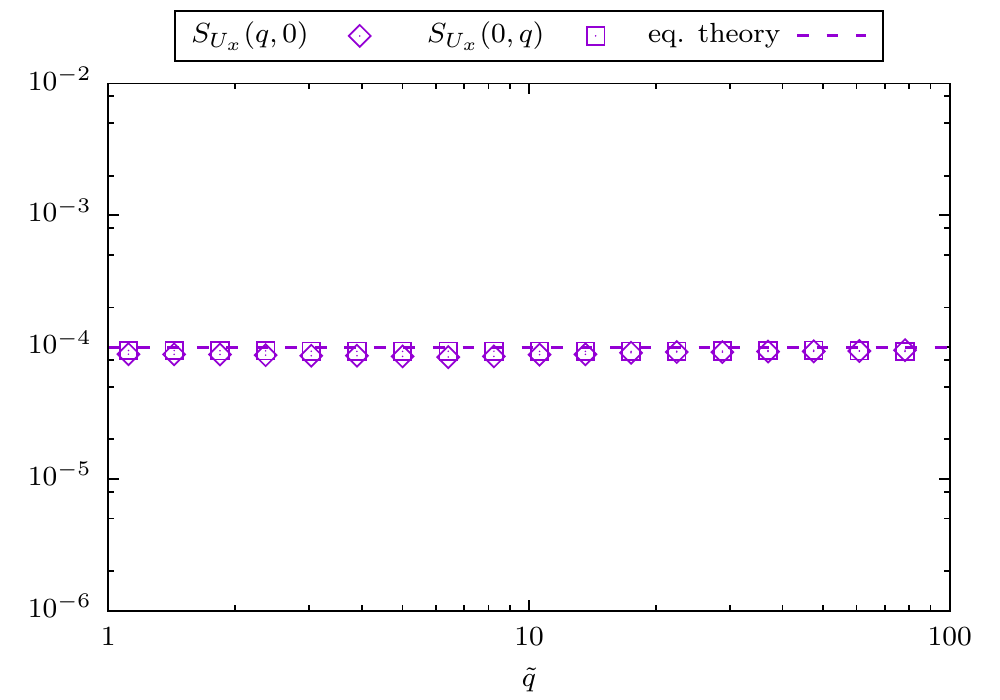}\\
\includegraphics[width=.6\columnwidth]{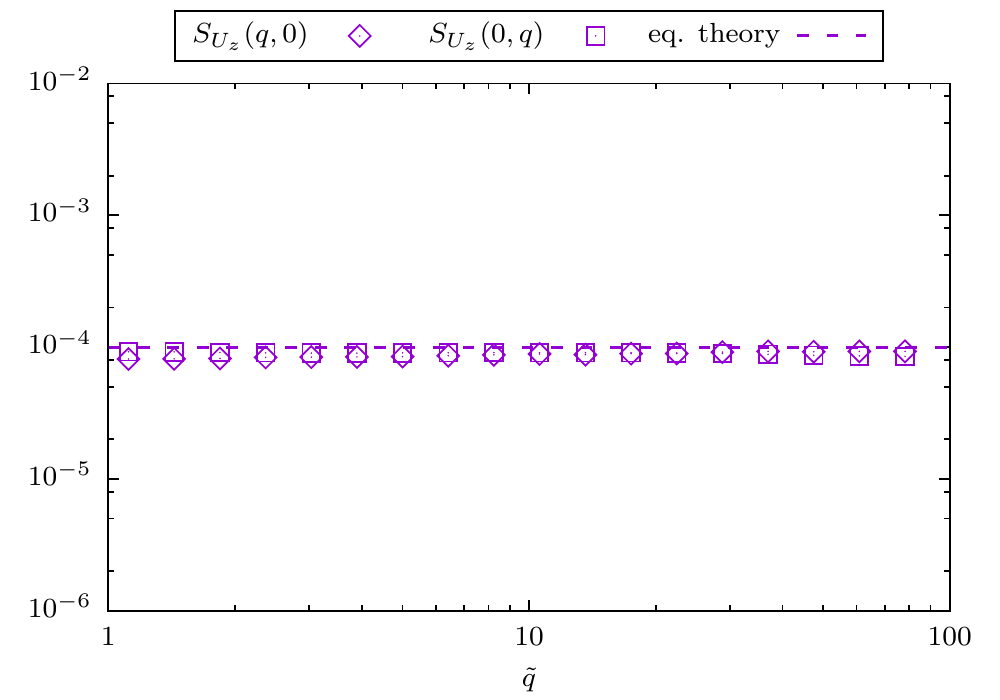}\\
\includegraphics[width=.6\columnwidth]{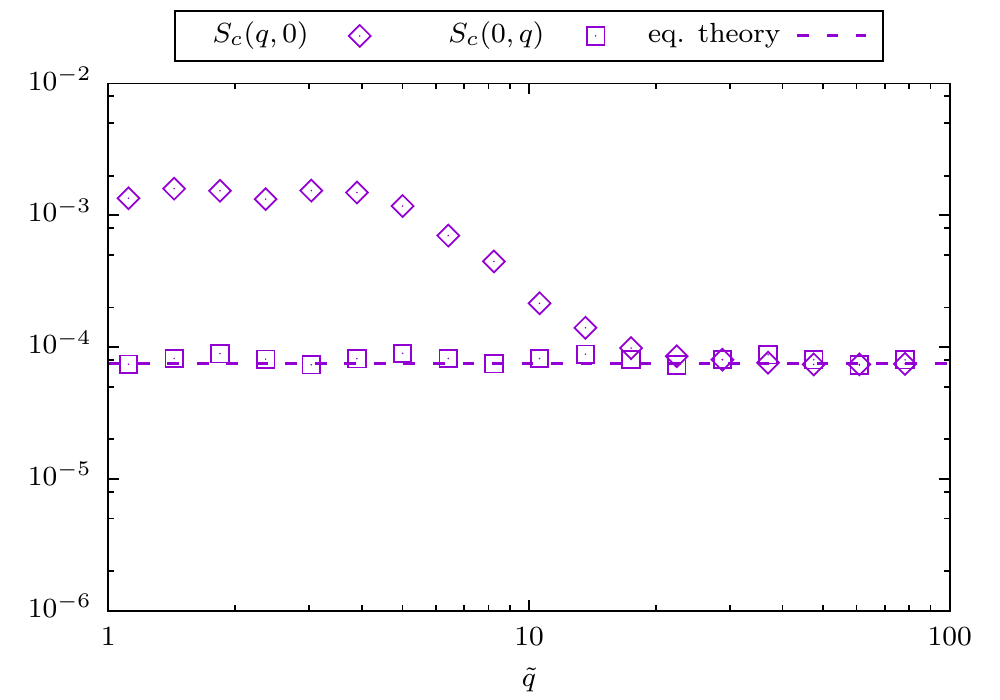}
\caption{Spectra of fluctuations of hydrodynamical fields around a NE background ($\nabla c_0\neq0$). Thermal energy and concentration gradient are fixed to $\kB T=10^{-4}$ and $\nabla c_0=0.005$, respectively. {\bf Top panel:} Velocity in the streamflow ($x$) direction; equilibrium theoretical prediction in Eq.~\eqref{eq:VELOCITY_EQ}. {\bf Central panel:} Velocity in the wall-to-wall ($z$) direction; equilibrium theoretical prediction in Eq.~\eqref{eq:VELOCITY_EQ}. {\bf Bottom panel:} Concentration fluctuations; equilibrium theoretical prediction in Eq.~\eqref{eq:CONCENTRATION_EQ}.\label{fig:spectrumNE}}
\end{figure}

\begin{figure}
\includegraphics[width=.6\columnwidth]{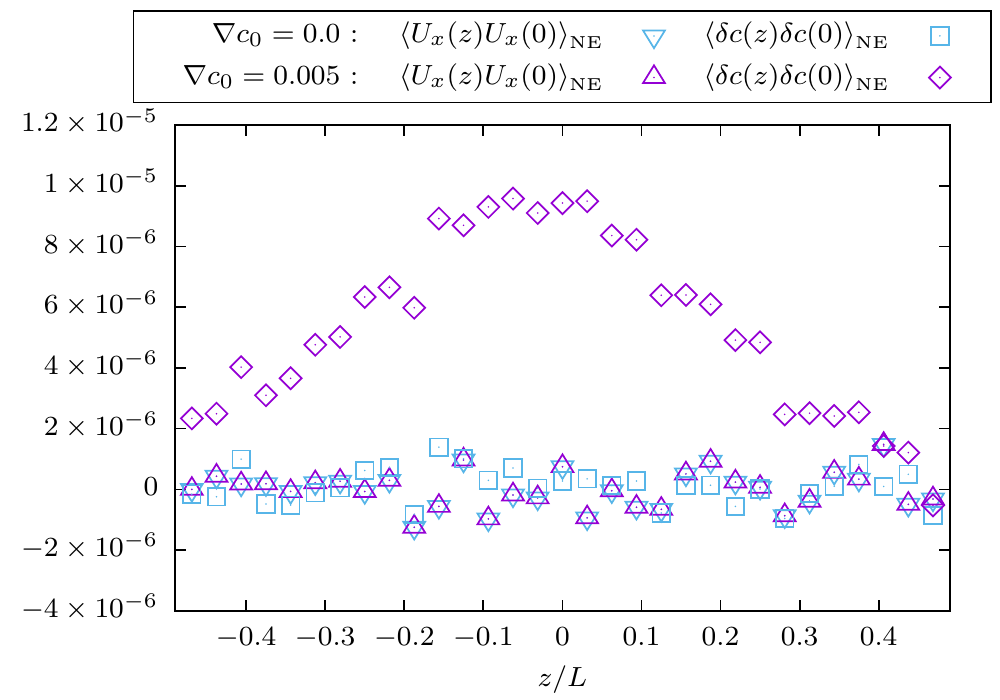}
\caption{Real space NE correlations of velocity and concentration fluctuations. They are obtained by subtracting to the measured correlations their respective equilibrium value (see Eqs.~\eqref{eq:VELOCITY_CORR}-\eqref{eq:CONCENTRATION_CORR}). Thermal energy is fixed to $\kB T=10^{-4}$.\label{fig:correlationrealspace}}
\end{figure}

\begin{figure}
\includegraphics[width=.6\columnwidth]{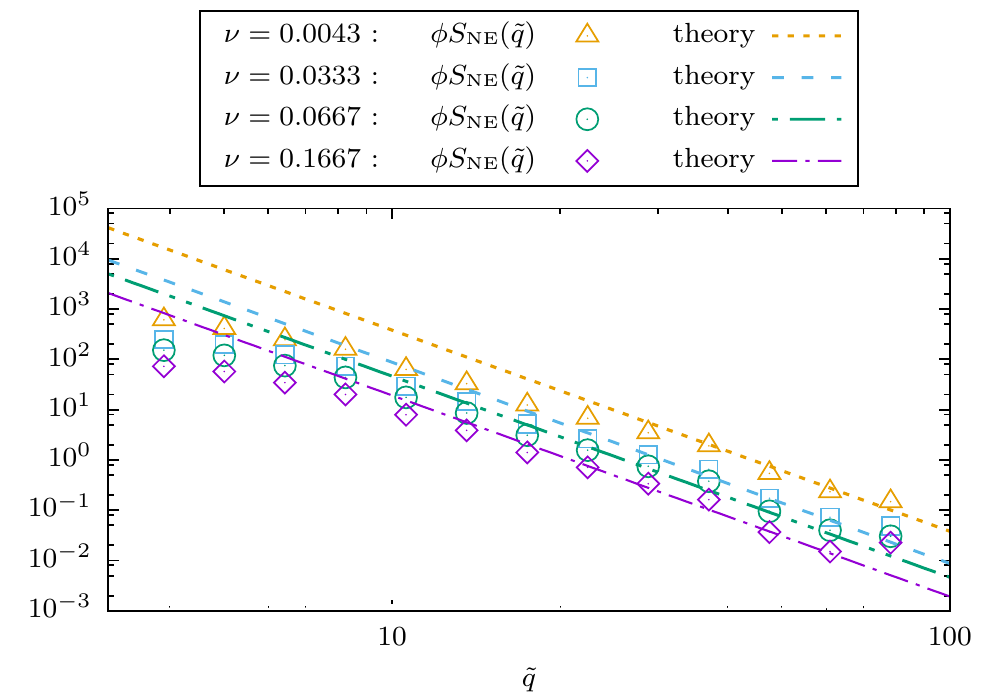}
\caption{Non-equilibrium structure factor contribution (see Eqs.~\eqref{eq:decompose}-\eqref{eq:prefactor}) as a function of the dimensionless wavenumber $\tilde{q}$ (see Eq.~\eqref{eq:qtilde}); the theoretical prediction, corresponding to $\tilde{S}_{\st{NE}}(\tilde{q})=\tilde{q}^{-4}$, is obtained from Eqs.~\eqref{eq:linearized1}-\eqref{eq:linearized3} for unbounded systems. We check the dependence on the kinematic viscosity $\nu$. Thermal energy, concentration gradient and diffusion coefficient are fixed to $\kB T=10^{-4}$, $\nabla c_0=0.01$ and $D=0.0043$, respectively.\label{fig:2}}
\end{figure}

\begin{figure}
\includegraphics[width=.6\columnwidth]{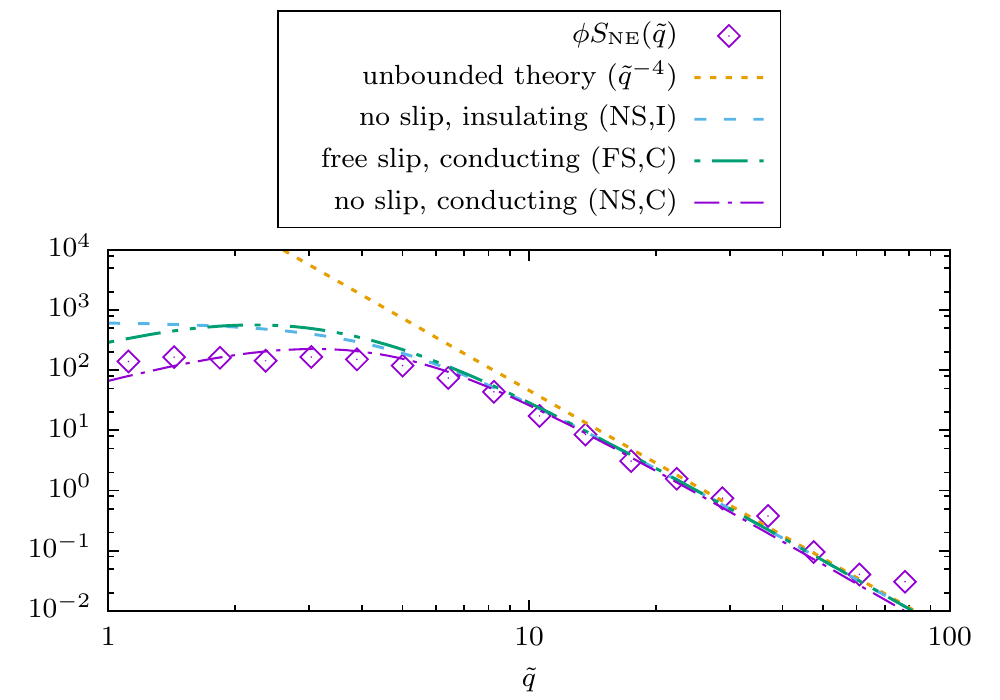}
\caption{Non-equilibrium structure factor contribution as a function of the dimensionless wavenumber $\tilde{q}$ (see Eq.~\eqref{eq:qtilde}). Different analytical formulas are checked, dependently on the boundary conditions \eqref{eq:BC}. Thermal energy, concentration gradient, diffusion coefficient and kinematic viscosity are fixed to $\kB T=10^{-4}$, $\nabla c_0=0.01$, $D=0.0043$ and $\nu=0.0667$, respectively.\label{fig:3}}
\end{figure}

\begin{figure}
\includegraphics[width=.6\columnwidth]{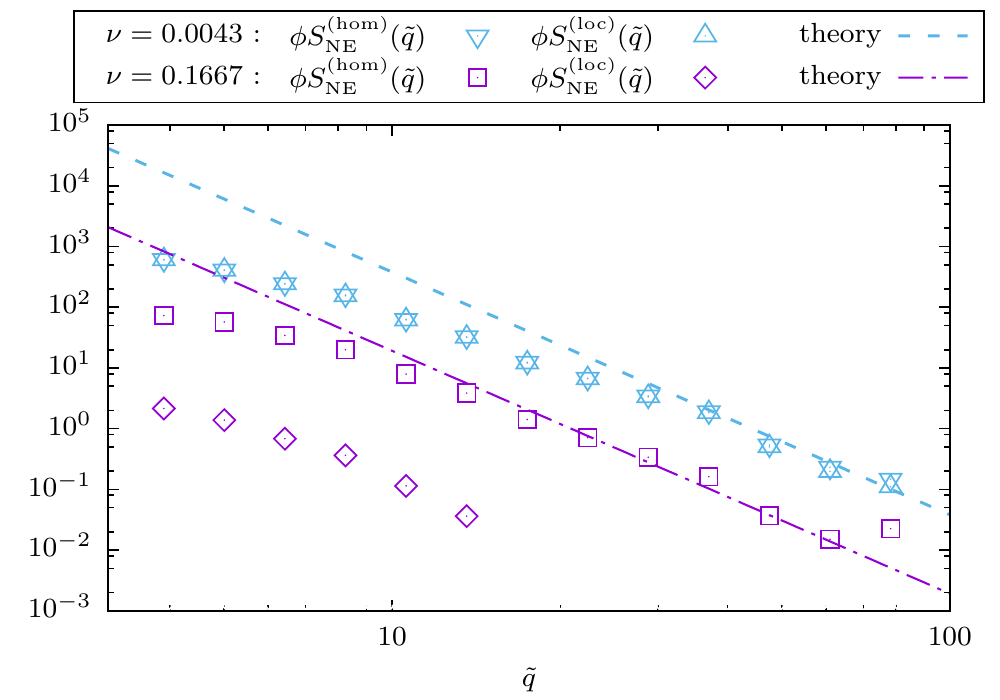}
\caption{Non-equilibrium structure factor contribution as a function of the dimensionless wavenumber $\tilde{q}$ (see Eq.~\eqref{eq:qtilde}). We compare the spatially homogeneous (hom) noise case with the local (loc) noise case, and check the dependence on the kinematic viscosity. Thermal energy, concentration gradient and diffusion coefficient are fixed to $\kB T=10^{-4}$, $\nabla c_0=0.01$ and $D=0.0043$, respectively.\label{fig:8}}
\end{figure}

In this section we start by describing the NE fluctuations. In Figure~\ref{fig:spectrumNE} we report results for the structure factors for the velocity and concentration fluctuations.
We observe that the structure factors for the velocities $U_{x,z}$ (top and central panels) are still homogeneous and isotropic in Fourier space and in agreement with the equilibrium prediction. For the fluctuations in the concentration field $\delta c$ (bottom panel), instead, the structure factors are anisotropic and mode-dependent. More specifically, they are well in agreement with the equilibrium prediction when $\qv=(0,q)$, while for $\qv=(q,0)$ we observe that the small-scale behaviour (large $\tilde{q}$, see Eq.~\eqref{eq:qtilde}) of the structure factor matches the equilibrium prediction, while it progressively overestimates this prediction at large scales (small $\tilde{q}$). When $\tilde{q} \approx 1$ this overestimate is about one order of magnitude. Correspondingly, the effect on the correlations in real space is highlighted in Figure~\ref{fig:correlationrealspace}: the two-point correlation function for the velocity (data shown only for the stream-flow velocity $U_x$) coincides with the delta-correlated equilibrium prediction~\eqref{eq:VELOCITY_CORR}, hence is (very) short-ranged; in contrast, the two-point correlation for the concentration highlights a correlation length that essentially spans the whole system size.
To characterize such NE fluctuations on a more quantitative basis, we therefore continue our analysis for the concentration field $c$ in a ``parallel flow approximation'', i.e. by taking the Fourier mode along the stream-flow direction ($\qv=(q,0)$). To facilitate a comparison with the existing literature on NE fluctuations we adopt the commonly used decomposition~\cite{ZarateBook}
\begin{equation}\label{eq:decompose}
S_c(q,0)={3}\frac{\kB T}{\bar{\rho}}\bar{c}_0(1-\bar{c}_0)\Square{1+ \phi \tilde{S}_{\st{NE}}(qL)},
\end{equation}
where
\begin{equation}\label{eq:prefactor}
\phi={\frac{1}{3}}\frac{L^4}{\bar{c}_0(1-\bar{c}_0)} \frac{(\nabla c_0)^2}{(\nu+D)D}.
\end{equation}
Starting from the data reported in Figure~\ref{fig:spectrumNE} and the decomposition~\eqref{eq:decompose}, we extracted the function $\phi\tilde{S}_{\st{NE}}(\tilde{q})$. The results are reported in Figure~\ref{fig:2}.
At small scales ($\tilde{q} \gg 1$) we observe the power-law scaling $\tilde{S}_{\st{NE}}(\tilde{q})\sim\tilde{q}^{-4}$. This is perfectly in agreement with the expected power-law behaviour $\tilde{S}_{\st{NE}}(\tilde{q})=\tilde{q}^{-4}$ predicted by the theory of NE fluctuations, which can be obtained from the equations of hydrodynamics linearized around a constant concentration gradient profile, i.e. Eqs.~\eqref{eq:linearized1}-\eqref{eq:linearized3}. We emphasize that we just used the decomposition~\eqref{eq:decompose} and added {\it no additional} fitting parameters. {We also} checked the goodness of the matching by changing the kinematic viscosity $\nu$, while keeping the diffusion coefficient $D$ unvaried. This can be done in the simulations thanks to the MRT generalization of the BGK model (see Eq.~\eqref{eq:MRT}), which allows to set different relaxation frequencies for different moments (see Eqs.~\eqref{eq:diff}-\eqref{eq:vis}). The plots reported in Figure~\ref{fig:2} show {changes} in agreement with the corresponding change of $\phi$ in Eq.~\eqref{eq:prefactor}. Thus, this result provides a very strong indication that the fluctuating LB methodology is quantitatively able to reproduce fluctuating hydrodynamics and the long-range spatial correlations peculiar of NE fluctuations~\cite{Kirkpatrick13,Kirkpatrick14,Kirkpatrick15,kirkpatrick2015,kirkpatrick2016,kirkpatrick2016b}. Going at smaller $\tilde{q}$, however, we observe in Figure~\ref{fig:2} that the power-law scaling $\sim\tilde{q}^{-4}$ becomes progressively underestimated by the numerically computed $\tilde{S}_{\st{NE}}(\tilde{q})$. This is attributed to finite-size effects induced by confinement. Indeed, due to the long-range nature of NE spatial correlations, NE structure factors are necessarily affected by the boundary conditions. There are various papers aimed at the quantitative characterization of $\tilde{S}_{\st{NE}}(\tilde{q})$ in presence of boundary conditions~\cite{Kirkpatrick15,ZarateSengers01,ZarateSengers02}. The results of these calculations share the common feature that the power-law behaviour $\sim \tilde{q}^{-4}$ is approached only at very small scales, i.e. $\tilde{S}_{\st{NE}}(\tilde{q}) \sim \tilde{q}^{-4}$ only for $\tilde{q} \rightarrow \infty$. {The} small-$\tilde{q}$ behaviour strongly depends on the boundary conditions used for both velocity and concentration. In what follows, we {discuss} three analytical (or semi-analytical) expressions for $\tilde{S}_{\st{NE}}(\tilde{q})$ that can be gathered from the literature. All of them treat the wall as impenetrable:
\be
\left.U_z\right|_{z=\pm L/2}=0.
\ee
This condition is strictly imposed in all the simulations performed. {One can then} impose either no-slip (NS) or free-slip (FS) boundary conditions for $U_x$, and independently either insulating (I) or conducting (C) boundary conditions for $\delta c$:
\be\label{eq:BC}
\begin{aligned}
\textup{(NS,I)} &: \left.(U_x,\pt_z \delta c)\right|_{z=\pm L/2} = \bm{0},\\
\textup{(FS,C)} &: \left.(\pt_zU_x,\delta c)\right|_{z=\pm L/2} = \bm{0},\\
\textup{(NS,C)} &: \left.(U_x,\delta c)\right|_{z=\pm L/2} = \bm{0},\\
\end{aligned}
\ee
The (NS,I) solution found in Eq. (30) of~\cite{Kirkpatrick15} is valid for $\nu\gg D$. For the (FS,C) boundary conditions one can get {an exact solution} (see Eq. (35) in~\cite{ZarateSengers01}). {Details} for (NS,C) are found in Eq. (20) of~\cite{ZarateSengers02} (see also (7.36) in~\cite{ZarateBook}). {In particular, this solution comes from a Galerkin truncation of exact equations} \footnote{This semi-analytical result underestimates by 20\% the exact large $\tilde{q}$ behavior, while it reproduces the small $\tilde{q}$ behavior within an error of 2\%~\cite{ZarateBook}}. The solution that better fits the data reported in Figure~\ref{fig:3} is (NS,C).
This is reasonable: we use a bounce-back for the kinetic population, thus reproducing the no-slip condition in the hydrodynamical limit; moreover, since fluctuations in the concentration are order $\kB T \ll 1$, one may also say that the conducting boundary condition fits well in those conditions where the concentration at the wall is much larger than $\kB T$, which is the case of the simulation data shown. However, we hasten to remark that a quantification of the boundary conditions with LB in presence of noise is currently missing in the literature. This surely stimulates further work in the future.\\
We finally studied the effect of the spatial dependence of the noise correlations in Eq.~\eqref{eq:noise-bin-expl}-\eqref{eq:noise-bin-expl-high}. Some authors before inspected the relative importance of mode-coupling effects and non-homogeneity in noise~\cite{Velasco91,ZarateSengers04}. In particular, they show that for a case with temperature (with thermal diffusivity $D_T$) the importance of the mode coupling effect with respect to the inhomogeneity in noise scales inversely proportional to the quantity $(\nu+D_T)D_T$. To verify this prediction, in Figure~\ref{fig:8} we reported the behavior of $\phi\tilde{S}_{\st{NE}}(\tilde{q})$ computed for two different simulations: one performed by implementing the noise according to \eqref{eq:noise-bin-expl}-\eqref{eq:noise-bin-expl-high} {and} using constant reference values for the mass densities (hom); the other by using the very same expression but using for the density their local (space-dependent) value (loc).
When both the transport coefficients ($\nu$ and $D$) are very small, the two simulations provide the same results. Instead, by increasing kinematic viscosity very different results are observed. This is a very preliminary result, and a quantitative study requires further numerical analysis, as well as a more precise connection with the results in~\cite{Velasco91,ZarateSengers04} when the role of $D_T$ is played by $D$.\\
Summarizing, the LB solver described in~\cite{belardinelli2015} generates a fluctuating hydrodynamical system that under the presence of a constant concentration gradient develops the typical long-range correlations characterizing NE fluctuations. Confinement effects also seem well captured, although a more systematic study of the boundary conditions emerging in the simulations is needed. Remarkably, neither fitting parameters nor corrective factors are needed to match numerics and analytical results.
%
\subsection{Non-Equilibrium Pressures}
\begin{figure}
\includegraphics[width=.6\columnwidth]{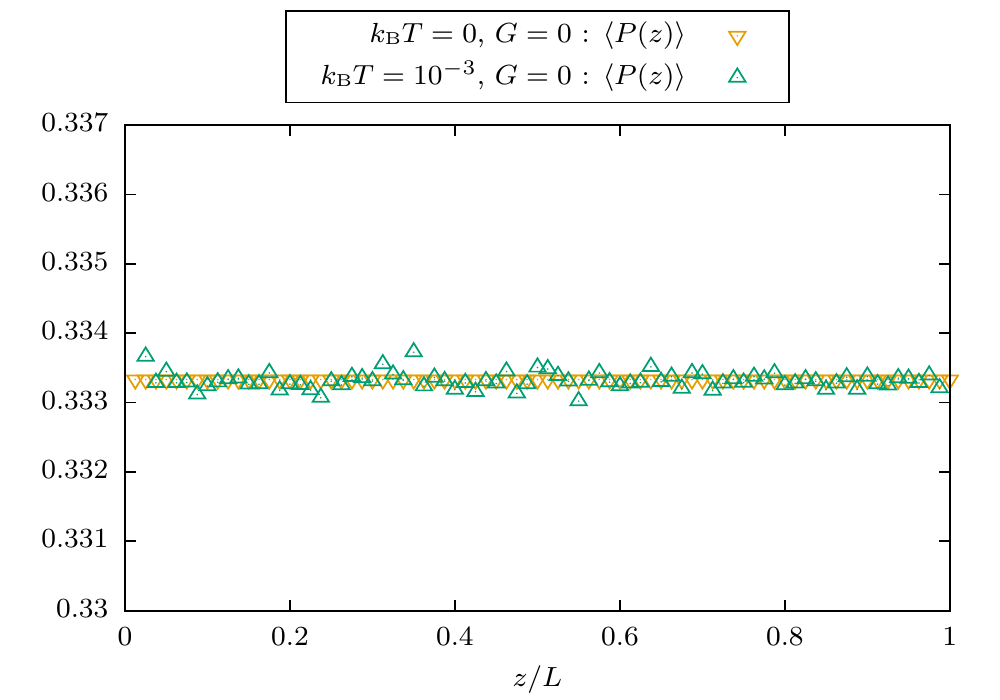}\\
\includegraphics[width=.6\columnwidth]{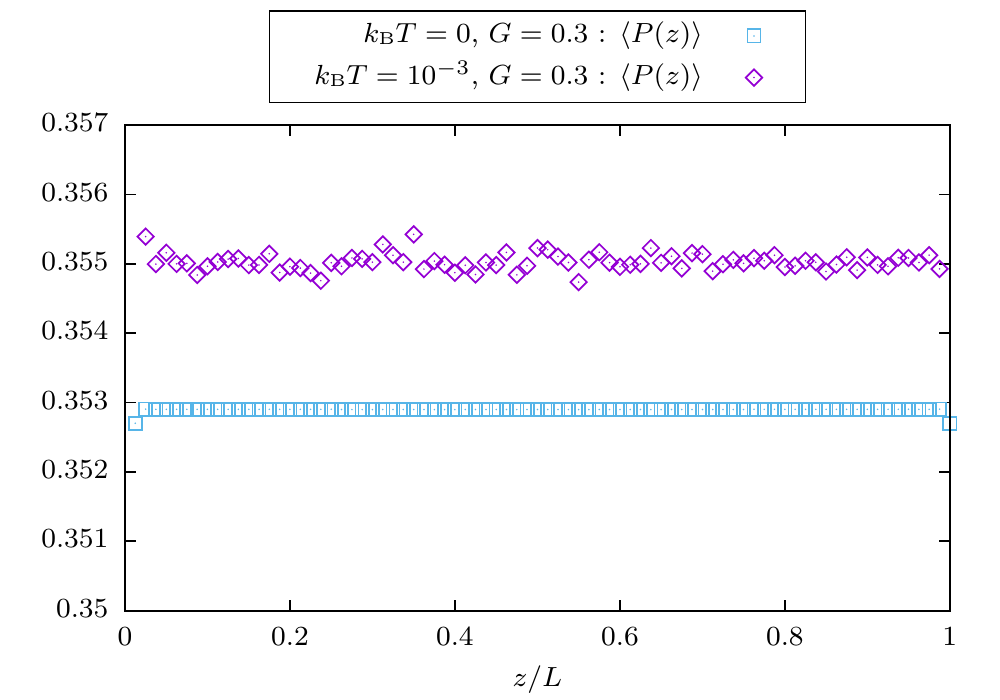}
\caption{Average \emph{total} pressure. Concentration gradient is fixed to $\nabla c_0=0.01$. {\bf Top panel:} Average total pressure for an ideal binary mixture ($G=0$). {\bf Bottom panel:} Average total pressure for a non-ideal binary mixture ($G>0$).\label{fig:pressureprofile}}
\end{figure}

\begin{figure}
\includegraphics[width=.6\columnwidth]{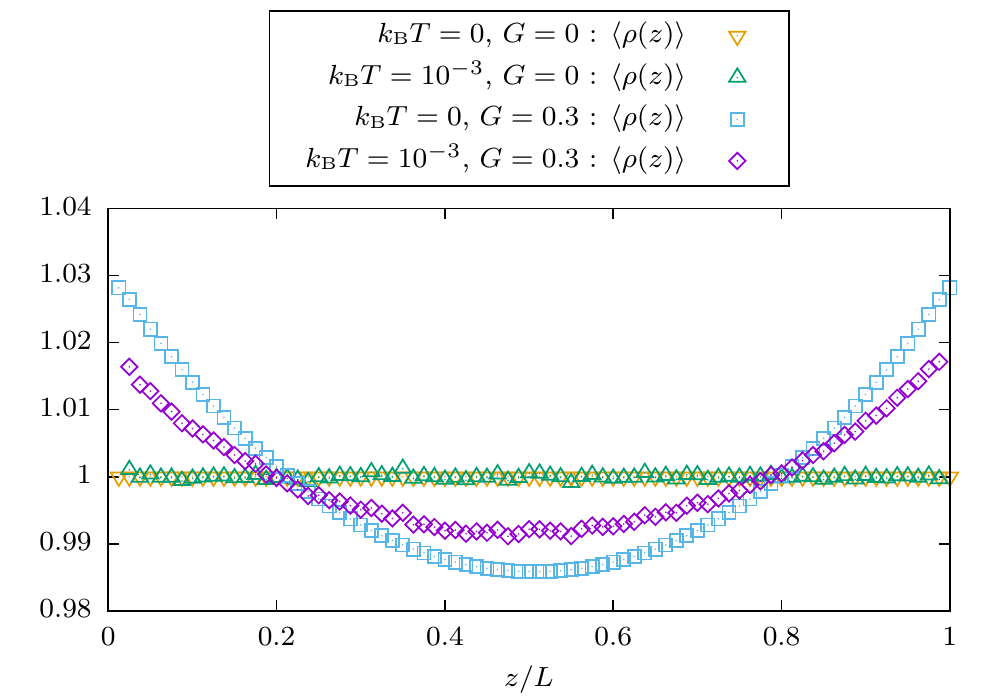}
\caption{Average total mass density in both the ideal ($G=0$) and non-ideal ($G>0$) cases. Concentration gradient is fixed to $\nabla c_0=0.01$.\label{fig:densityprofile}}
\end{figure}

\begin{figure}
\includegraphics[width=.6\columnwidth]{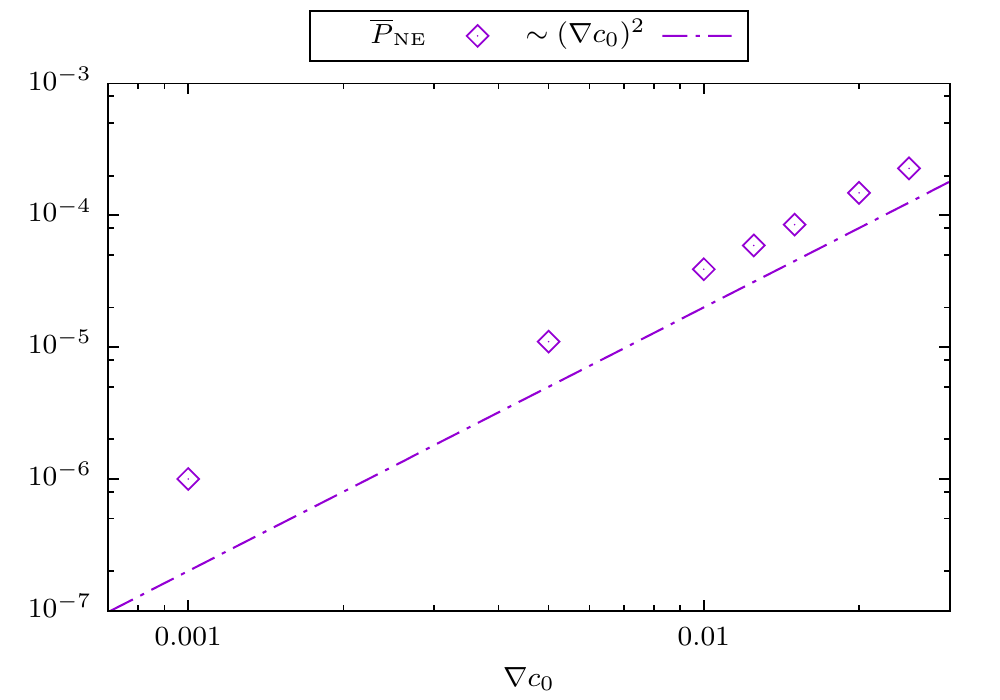}
\includegraphics[width=.6\columnwidth]{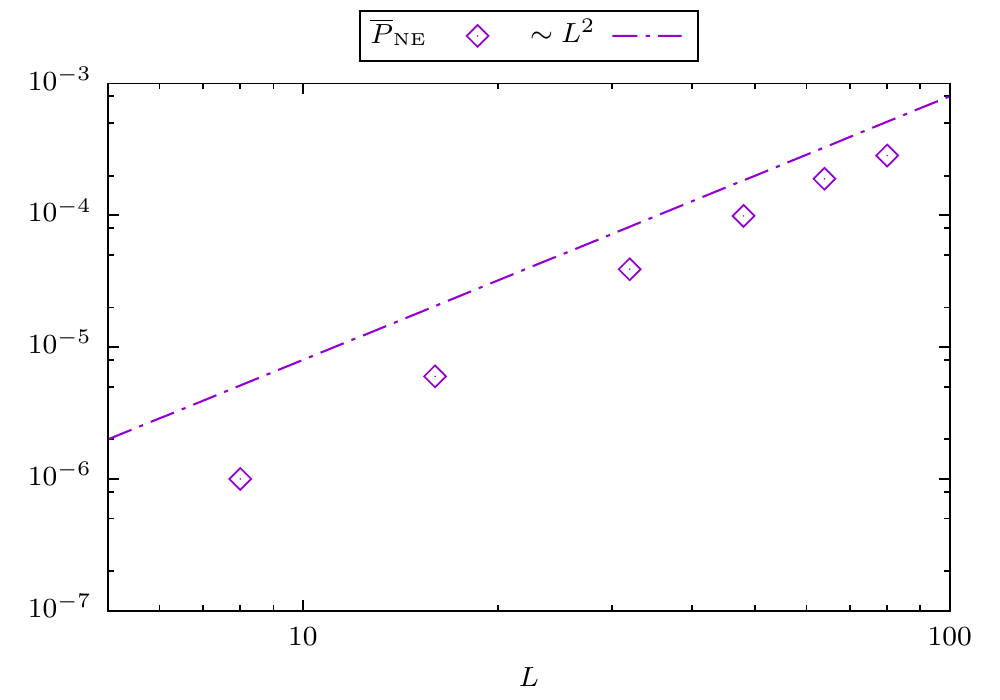}
\caption{Scaling laws for the NE contribution to the \emph{spatial} average pressure computed according to Eq.~\eqref{eq:NEQP}. Thermal energy is fixed at $\kB T=10^{-4}$. {\bf Top panel:} NE pressure contribution at fixed wall-to-wall separation as a function of the concentration gradient. Wall-to-wall separation in fixed to $L=32$ grid points. {\bf Bottom panel:} NE pressure contribution at fixed concentration gradient as a function of the wall-to-wall separation. Concentration gradient is fixed to $\nabla c_0=0.01$.\label{fig:4}}
\end{figure}

Recent papers of the literature~\cite{Kirkpatrick13,Kirkpatrick14,Kirkpatrick15,kirkpatrick2015,kirkpatrick2016,kirkpatrick2016b} supported the idea that the long-range effects deriving from NE fluctuations (see Figure~\ref{fig:correlationrealspace}) cause a NE ``Casimir'' pressure. The rationale behind this effect hinges on the connection between the pressure and concentration fluctuations. In a nutshell, the \emph{local} equilibrium assumption relates mass density and concentration to pressure through an equation of state $P=P(\rho,c)$ satisfying $\grad P=\bm{0}$ (see equation \eqref{eq:EQUATIONofSTATE} for the case at hand), which is expected to be still valid in average. Fluctuations of $\rho$ and $c$ are then accompanied by fluctuations of $P$ that are vanishing at linear order. By keeping the first non vanishing terms, one gets~\cite{kirkpatrick2015,kirkpatrick2016b}
\be\label{eq:NEQpressure}
P_{\st{NE}}(z)=\frac{1}{2}A_c \Angle{|\delta c(z)|^2}_{\st{NE}},
\ee
where the vanishing of the linear order is used to express $\delta\rho$ in terms of $\delta c$. The constant $A_c$ plays the role of a second order coefficient, and is a function of the background fields computed at their respective reference values. Two important comments are in order. {\it First}, based on the prediction for the NE pressure in~\eqref{eq:NEQpressure}, one would expect NE Casimir pressures to be triggered by the non-ideality of the mixture (see Eq.~\eqref{eq:EQUATIONofSTATE}): an ideal equation of state ($G=0$) would just deliver $A_c=0$ and hence $P_{\st{NE}}=0$. {\it Second,} the NE correlation $\Angle{|\delta c(z)|^2}_{\st{NE}}$ may be non homogeneous in space, depending on the choice of the boundary conditions~\cite{Kirkpatrick15,kirkpatrick2015,kirkpatrick2016b}. Thus, the resulting NE pressure in~\eqref{eq:NEQpressure} is space-dependent and one may wander how this could be reconciled with an average mechanical balance. Indeed, the mode coupling effect triggers NE effects only in the concentration fluctuations, while velocity fluctuations are unchanged (see Figure~\ref{fig:spectrumNE}); thus, one would expect the equilibrium condition of a constant (average) pressure to be recovered. As already pointed out~\cite{kirkpatrick2016b}, the mechanism of compensation is a NE renormalization of the background profile which provides a zero derivative of the {\it total} pressure. In other words, the pressure may be seen as the sum of an equilibrium contribution and the NE contribution of Eq.~\eqref{eq:NEQpressure}; $z$-dependency in the latter causes the former to be $z$-dependent in such a way that the total pressure has zero derivative.\\ 
Based on the numerical model that we used, we are in a condition to test directly these properties. {In practice, {\it total} pressure is evaluated by its mechanical definition, that is as half the trace of the pressure tensor~\cite{sbragaglia2013interaction}, whose bulk behaviour is expected to coincide with \eqref{eq:EQUATIONofSTATE} in the hydrodynamical limit.} Results are reported in Figures~\ref{fig:pressureprofile}-\ref{fig:densityprofile} and fully confirm the above views.
Specifically, we fixed a non-zero concentration gradient $\nabla c_0=0.01$, and we performed simulations with $\kB T>0$ and $\kB T=0$ for an ideal mixture ($G=0$, Figure~\ref{fig:pressureprofile}, top panel) and for a non-ideal mixture ($G>0$, Figure~\ref{fig:pressureprofile}, bottom panel). We observed that the {\it total} pressure profiles are homogeneous in $z$ and that the pressure receives a correction by thermal fluctuations {\it only when} $G>0$. Only when the pressure receives a correction, the average density profile slightly changes with thermal fluctuations (Figure~\ref{fig:densityprofile}).\\
These facts said, we wanted to further characterize the NE Casimir pressure from our simulations, hence we sticked with a non-ideal mixture with fixed $G>0$. The \emph{spatial} average pressure will then depend on $L$, $\kB T$ and $\nabla c_0$, i.e. $\overline{P}=\overline{P}(L,\kB T,\nabla c_0)$. To make progress we wanted to study the scaling properties of the NE Casimir pressure as a function of the system size $L$ and concentration gradient $\nabla c_0$. We emphasize that fluctuations are expected to induce pressure effects also in equilibrium conditions ($\nabla c_0=0$), and that the latter effects are particularly large close to the critical point (critical ``Casimir'' pressure) and decay to zero at large $L$~\cite{gambassi2009}. For the parameters chosen~\cite{CHEM09} the critical point corresponds to $G=2$, while we kept $G=0.3$ in all the non-ideal simulations. In such conditions thermal fluctuations only trigger some small effects in equilibrium conditions, that we detect only at the smallest $L$ considered; however, aiming at characterizing the NE pressure at changing $L$, we needed to remove such small contributions. We proceeded as follows. For a given system size $L$, we first performed a numerical simulation in equilibrium conditions ($\nabla c_0=0$) without thermal fluctuation ($\kB T=0$); we then repeated the simulation with the desired $\kB T$. In both simulations we have computed the average pressure and we estimated the pressure difference induced by thermal fluctuations as
\begin{equation}\label{eq:EQP}
\Delta P^{\eq}(L,\kB T)=\overline{P}(L,\kB T,0)-\overline{P}(L,0,0).
\end{equation}
Then, for the desired $\nabla c_0 >0$, we performed two other simulations without thermal fluctuations ($\kB T=0$) and with the desired $\kB T$. The NE contribution to the spatial average pressure has been identified as
\begin{equation}\label{eq:NEQP}
\overline{P}_{\st{NE}}(L,\kB T,\nabla c_0) = \overline{P}(L,\kB T,\nabla c_0)-\overline{P}(L,0,\nabla c_0) -\Delta P^{\eq}(L,\kB T).
\end{equation}
In Figure~\ref{fig:4} we plot the measured $\overline{P}_{\st{NE}}$ as a function of $\nabla c_0$ and $L$. 
While the scaling $\overline{P}_{\st{NE}}\sim(\nabla c_0)^2$ is in agreement with the theoretical predictions~\cite{Kirkpatrick13,Kirkpatrick14,Kirkpatrick15,kirkpatrick2015,kirkpatrick2016,kirkpatrick2016b}, the behavior of $\overline{P}_{\st{NE}}\sim L^2$ reflects the two dimensional character of the system. This can be understood by looking at the unbounded behavior $\sim |\qv|^{-4}$ of $\Angle{|\delta c(\qv)|^2}_{\st{NE}}$ in Fourier space. Indeed, the computation of the NE average pressure from Eq.~\eqref{eq:NEQpressure} requires the integration of $|\qv|^{-3}$ in a two dimensional system, in contrast with the integration of $|\qv|^{-2}$ for a three dimensional system, as those considered in~\cite{Kirkpatrick13,Kirkpatrick14,Kirkpatrick15,kirkpatrick2015,kirkpatrick2016,kirkpatrick2016b}. Consequently, if an infrared cutoff proportional to $L^{-1}$ is introduced, then a two dimensional system furnishes $\sim L^2$, while a three dimensional system gives $\sim L$. Predicting the offsets requires the complete control of the boundary conditions.
\section{Conclusions\label{sec:Conclusions}}
We applied the fluctuating lattice Boltzmann (LB) methodology described in~\cite{belardinelli2015} to a system out of thermodynamic equilibrium. Specifically, we considered a binary mixture confined between two parallel walls in presence of a constant concentration gradient in the wall-to-wall direction. We studied structure factors and spatial correlations of the velocity and concentration fluctuations, and found good agreement with the theoretical predictions of fluctuating hydrodynamics~\cite{ZarateBook}. We further inspected the behavior of the resulting NE pressure as a function of both the concentration gradient and the wall-to-wall distance, and verified the correctness of the corresponding expected scaling laws~\cite{Kirkpatrick15,kirkpatrick2015,kirkpatrick2016b}, in agreement with a constant average total pressure. The results here reported naturally warrant other future quantitative studies in the context of LB methodology. Specifically, the analysis of the structure factors revealed the necessity of a better control in implementing the boundary conditions in presence of thermal fluctuations. Furthermore, the extension of the Chapman-Enskog procedure to the fluctuating case is missing. In this sense, the results of this paper support the convergence of fluctuating LB towards fluctuating hydrodynamics.\\
On a more general perspective, we remark that NE effects are continuously invoked in a variety of situations of experimental interest involving complex hydrodynamics. These include studies with colloidal suspensions~\cite{croccolo2006,oprisan2010,giavazzi2016,oprisan2017}, transient and enhanced diffusion effects \cite{Donev11,donev2014,cerbino2015,baaske2016}, driven active matter~\cite{kirkpatrick2018}, complex polymeric fluids~\cite{samanta2017}, finite Reynolds numbers fluids~\cite{Zarate2018}. In particular, for the future, it could be insightful to design experiments involving colloidal particles exhibiting a mechanical-chemical coupling with the fluid~\cite{Karpral18}, in such a way that NE fluctuations effects can be indirectly reconstructed and studied from the particles trajectories. The LB methodology has proven capable of remarkable versatility in the simulation of colloidal particles~\cite{DunwegReview,Aidun10,Zhang11,Kangetal14,Schilleretal18}, hence results of the present paper are instrumental for the use of LB as a validated methodology to support and complement experimental studies in the aforementioned direction.

\appendix*
\section{Definitions of structure factors}
In this appendix we report the essential technical details for the computations of the structure factors of a generic scalar field $\varphi=U_x,U_z,c$. Given the wave vector $\qv=(q_x,q_z)$, we started from the partial Fourier transform around a generic $z$-dependent background $\varphi_0(z)$ 
\begin{equation}\label{eq:SF1}
\delta \varphi(q_x,z,t)= \frac{1}{\sqrt{L_x}}\int_{0}^{L_x}\d x (\varphi(x,z,t)-\varphi_0(z)) e^{-iq_x x}.
\end{equation}
Based on Eq.~(6.30) in~\cite{ZarateBook}, we defined the quantity ${\cal C}_\varphi(q_x,z,z')$ through the equal-time mixed correlation:
\begin{equation}
\Angle{\delta \varphi(q_x,z,t)^*\delta \varphi(q_x',z',t)} = {\cal C}_\varphi(q_x,z,z')2\pi\delta(q_x-q_x'),
\end{equation}
where $\Angle{\dots}$ indicates the ensemble average computed via the equal time average in the statistically stationary state. We then Fourier-transformed in $z$ and $z'$ to define the structure factor (see Eq.~(31) in~\cite{ZarateSengers01}):
\begin{equation}\label{eq:integral}
S_\varphi(\qv)=\frac{1}{L}\int_{-L/2}^{+L/2}\d z\d z' e^{-i q_z (z-z')}  {\cal C}_\varphi(q_x,z,z').
\end{equation}
We can also write
\begin{equation}
\Angle{\delta \varphi(\qv,t)^*\delta \varphi(\qv',t)} = S_\varphi(\qv)(2\pi)^2\delta(\qv-\qv'),
\end{equation}
which gives $S_\varphi(\qv)=\Angle{|\delta \varphi(\qv,t)|^2}$ on a two dimensional lattice, where $(2\pi)^2\delta(\qv-\qv')$ is replaced by $\delta_{\qv,\qv'}$.

\bibliographystyle{apsrev4-1}
\bibliography{Bib}

\end{document}